\def\mtiny{\vrule width 0pt}
\def\mrm#1{\mathrm{#1}}
\def\DZ{\relax\ifmmode{D^0}\else{$\mrm{D}^{\mrm{0}}$}\fi}
\def\KZ{\relax\ifmmode{K^0}\else{$\mrm{K}^{\mrm{0}}$}\fi}
\def\BZ{\relax\ifmmode{B^0}\else{$\mrm{B}^{\mrm{0}}$}\fi}
\def\DZB{\relax\ifmmode{\overline{D}\mtiny^0}
        \else{$\overline{\mrm{D}}\mtiny^{\mrm{0}}$}\fi}
\def\KZB{\relax\ifmmode{\overline{K}\mtiny^0}
        \else{$\overline{\mrm{K}}\mtiny^{\mrm{0}}$}\fi}
\def\BZB{\relax\ifmmode{\overline{B}\mtiny^0}
        \else{$\overline{\mrm{B}}\mtiny^{\mrm{0}}$}\fi}

\providecommand{\KPI} {\mbox{$\mathrm{K^-}\pi^+ $}}
\providecommand{\WS} {\mbox{$\mathrm{K^+}\pi^- $}}

\documentclass{ws-p8-50x6-00}

\begin{document}

\title{$D^0$ Mixing and Cabibbo Suppressed Decays\hfill}

\author{D. M. Asner}

\address{Department of Physics, University of California,\\
Santa Barbara 93106-9530, USA} 


\maketitle\abstracts{The CLEO collaboration reported observation of the `wrong sign' decay
$D^0\!\to\!K^+\pi^-$ in 1993.  Upgrades
have been made to the CLEO detector\cite{CLEO}, including installation
of a silicon vertex detector\cite{SVX}, which provide substantial improvements
in sensitivity to $D^0\!\!\to\!K^+\pi^-$.
The vertex detector enables the reconstruction of the
proper lifetime\cite{D_life} of the $D^0$, and so provides sensitivity
to $D^0\!\!-\!\overline{D}\vphantom{D}^0$ mixing. We will
give preliminary results on the rate of `wrong sign' decay 
and $D^0\!\!-\!\overline{D}\vphantom{D}^0$ mixing using data from the
$9.1\,$fb$^{-1}$ of integrated luminosity
that has been accumulated with the upgrades in place.
In addition, we will give sensitivity estimates
of on-shell $D^0\!\!-\!\overline{D}\vphantom{D}^0$ mixing
derived from measurement of the lifetime measured
with decays of the $D^0$ to $CP$ eigenstates such as
$K^+K^-$, $\pi^+\pi^-$, and $K_{\rm S}\phi$.
}

\section{Introduction}
Ground state mesons such as the $\KZ$, $\DZ$, and $\BZ$,
which are electrically neutral and contain a quark and
antiquark of different flavor, can evolve into their respective
antiparticles, the $\KZB$, $\DZB$, and $\BZB$.
The rate measurements of $\KZ\!-\!\KZB$ mixing and
$\BZ\!-\!\BZB$ mixing have guided both the elucidation
of the structure of the Standard Model and the determination
of the parameters that populate it.  These mixing measurements
permit crude, but accurate, estimates of the 
masses of the
charm and top quark masses prior to direct observation of those
quarks at the high energy frontier.

Within the framework of the Standard Model the evolution of
a $\DZ$ into a $\DZB$ is expected to be infrequent, for two reasons.
First, the overall $\DZ$ decay amplitude is not Cabibbo suppressed,
in distinction to the $\KZ$ and $\BZ$ cases. In all cases the
mixing amplitude is (at least) double Cabibbo suppressed; consequently,
the magnitudes of $x$ and $y$, which are the ratios of the mixing
amplitude via virtual and real intermediate states, respectively,
to the mean decay amplitude, are not expected to exceed
$\tan^2\theta_c\approx0.05$ for $\DZ\!-\!\DZB$ mixing.\cite{Wo}
\begin{eqnarray}
\begin{array}{rcl}
x & = & \displaystyle{\Delta M\over\overline{\Gamma}}\;\;\;\;\;\;\; y = \displaystyle{\Delta \Gamma\over2\overline{\Gamma}}\\ \nonumber
\end{array}
\end{eqnarray}
Three out of four of the analogous
ratios for the $\KZ$ and $\BZ$ systems have been measured and are
all close to unity.  Second, the near degeneracy in mass of the $d$ and $s$ 
quarks relative to the $W$ boson causes the 
Glashow-Illiopolous-Maini (GIM) cancellation
to be particularly effective\cite{DaKu}. This drives the relative $\DZ$
amplitudes down by a rather uncertain additional factor of 10 to $10^3$.
It was the \emph{absence} of perfect GIM cancellation that 
permitted the inference of crude values of $m_c$ and $m_t$ 
from the various measurements
of $\KZ$ and $\BZ$ mixing, prior to the direct observation of
the $c$ and $t$ quarks.

The observation of a value of $|x|$ in the $\DZ\!-\!\DZB$ system in 
excess of about $5\times 10^{-3}$ might  be evidence
of incomplete GIM-type cancellations among new families of particles,
such as supersymmetric partners of quarks.\cite{Coea}  The evidence would
be most compelling if either the mixing amplitude exhibited a large CP
violation, or if the Standard Model contributions could be decisively
determined.  It is possible that in the Standard Model that
$|y|>|x|$,\cite{GoPe} and a determination of $y$ allows the estimation of
at least some of the long-distance Standard Model contributions
to $x$.

The Standard Model predicts that $\DZ\!-\!\DZB$ mixing
is likely to proceed through real intermediate states and will cause
the decays to CP+ final states to have the shorter lifetime.  This
situation would cause \emph{constructive} interference between
mixing and decay in the process $\DZ\to K^+\pi^-$.

\begin{figure}[h]
\parbox{.24\textwidth}{a)}\parbox{.24\textwidth}{b)}\parbox{.24\textwidth}{c)}\parbox{.24\textwidth}{d)}
\parbox{.24\textwidth}{\epsfig{figure=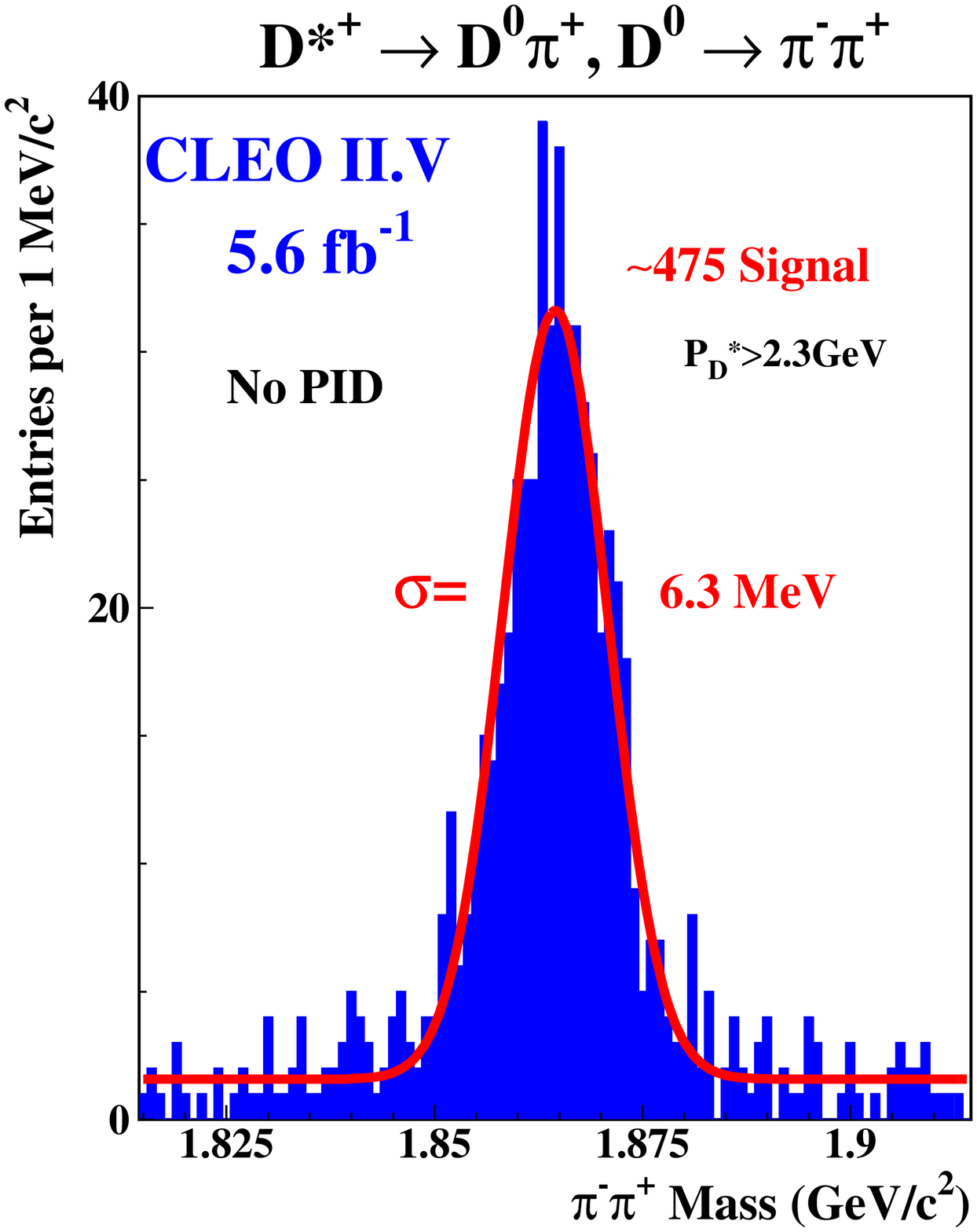,width=.24\textwidth,height=2.0in,
clip=, bbllx=42, bblly=161, bburx=552, bbury=736}}
\parbox{.24\textwidth}{\epsfig{figure=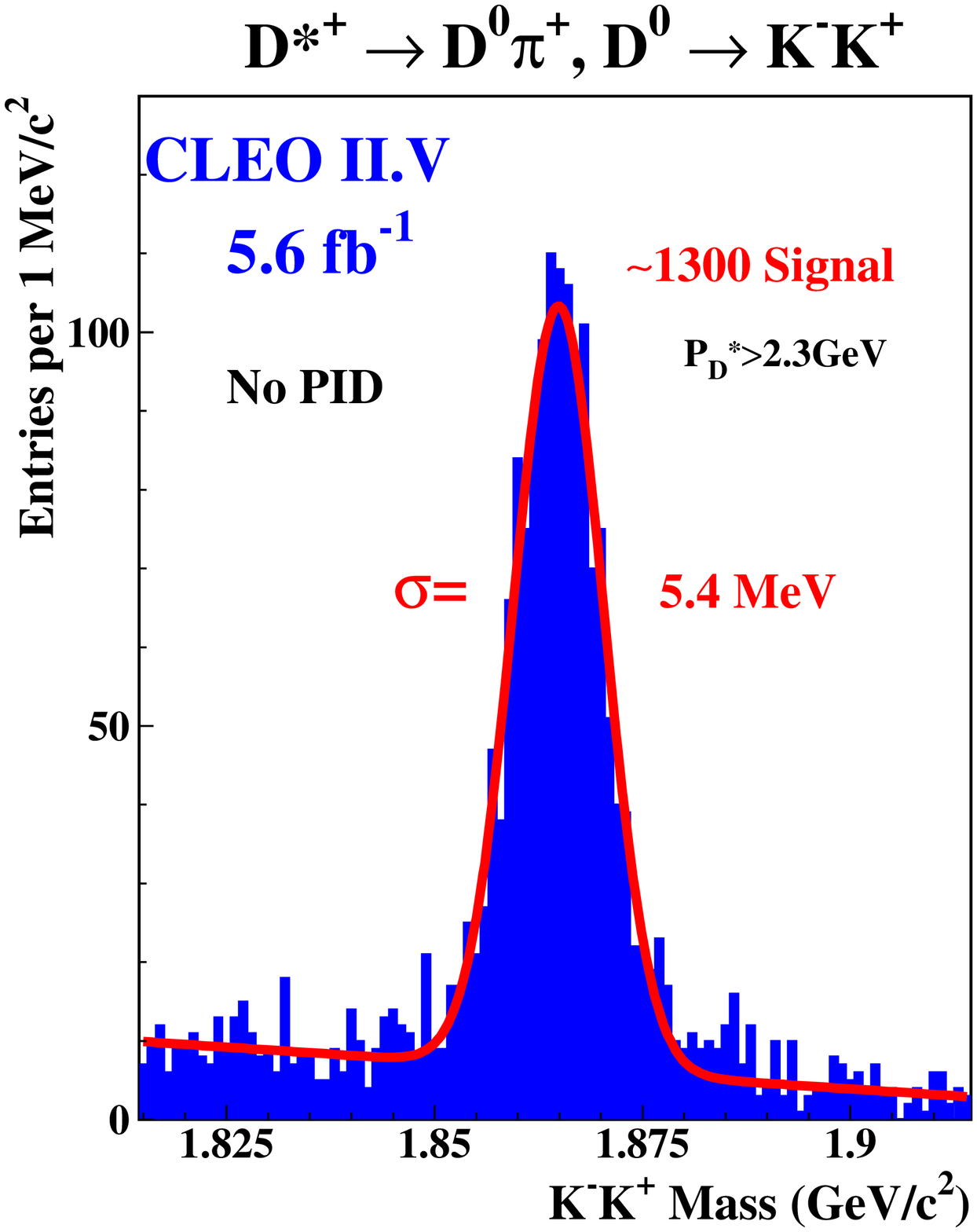,width=.24\textwidth,height=2.0in,
clip=, bbllx=42, bblly=161, bburx=552, bbury=736}}
\parbox{.24\textwidth}{\epsfig{figure=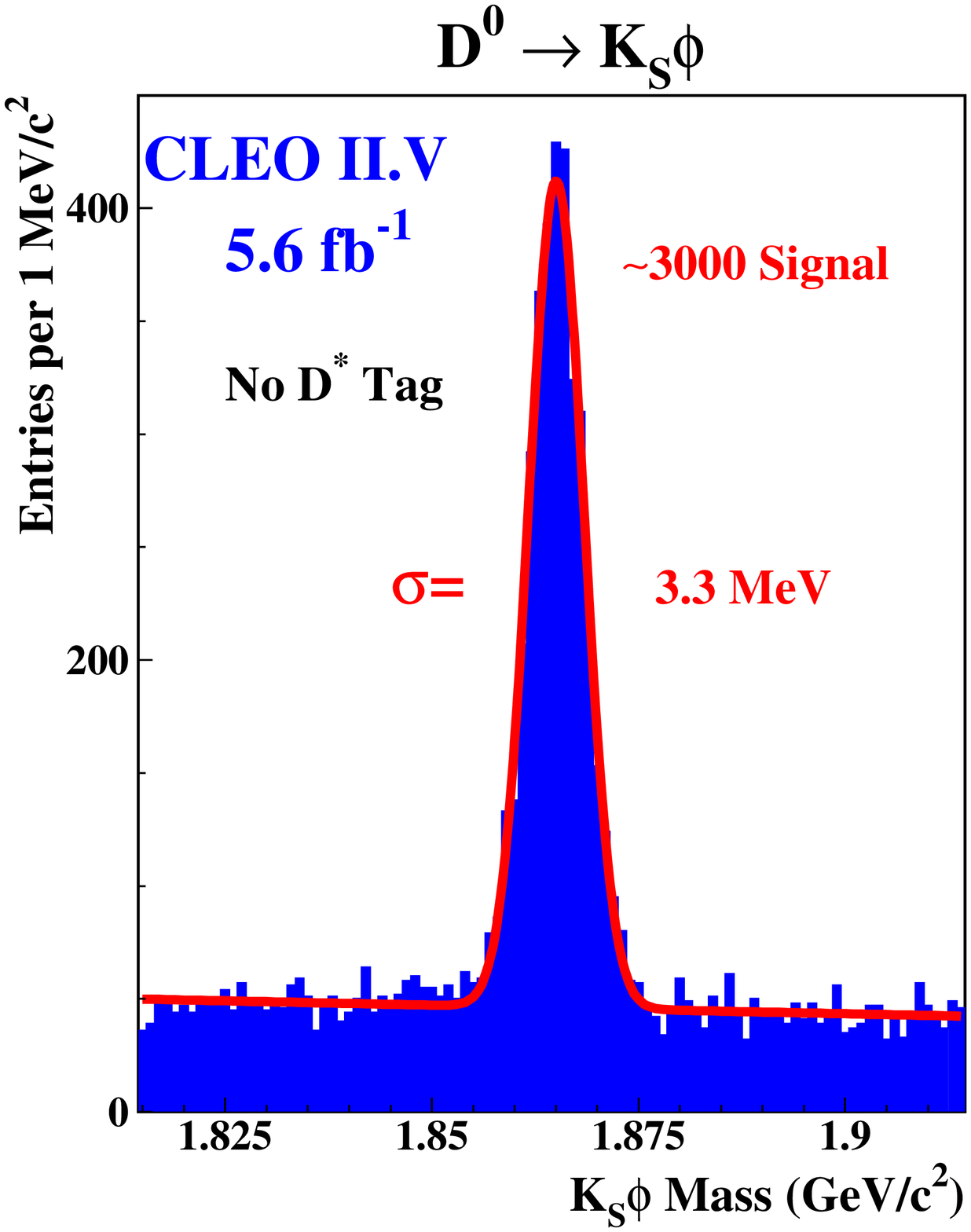,width=.24\textwidth,height=2.0in,
clip=, bbllx=42, bblly=161, bburx=552, bbury=736}}
\parbox{.24\textwidth}{\epsfig{figure=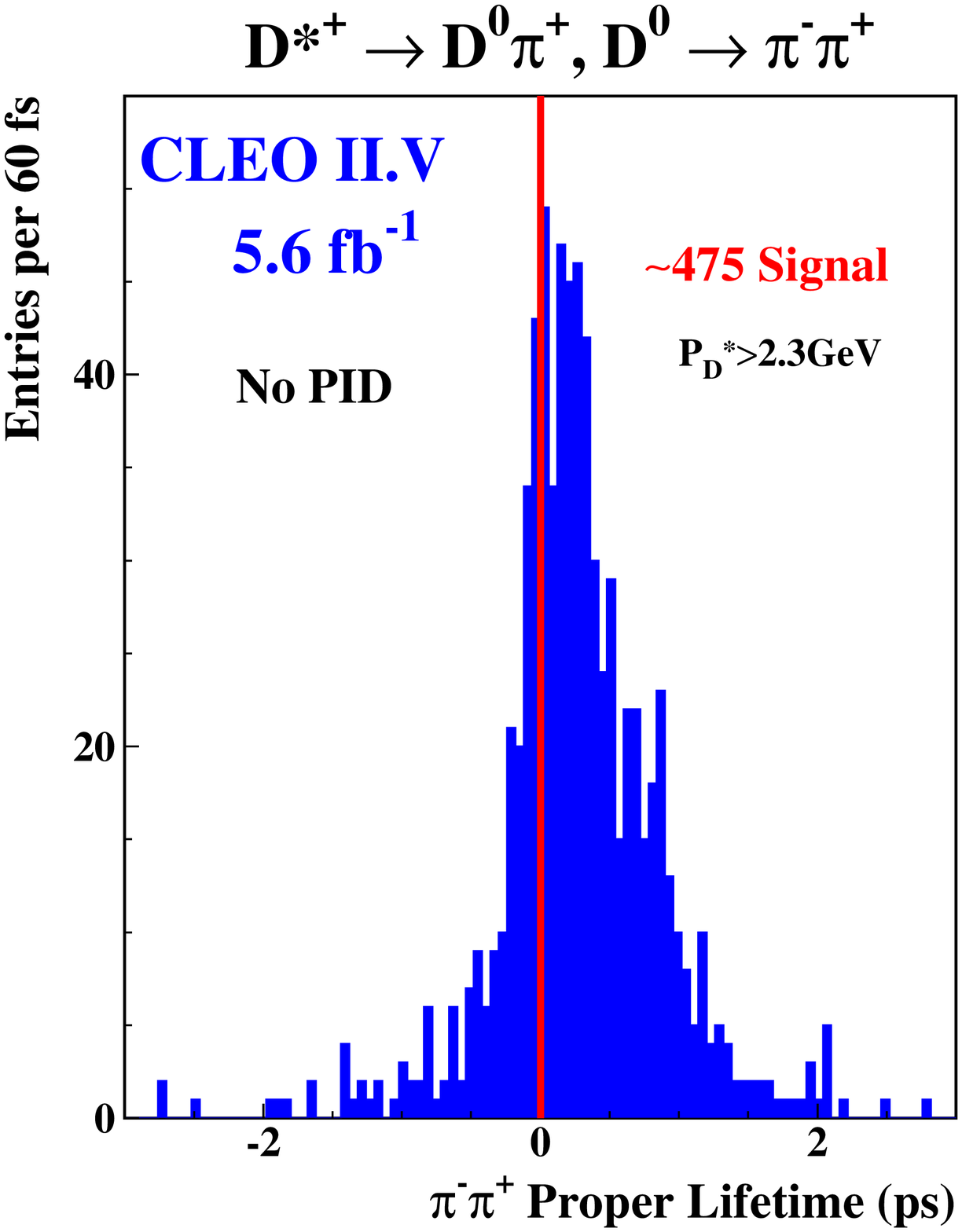,width=.24\textwidth,
height=1.in,clip=, bbllx=42, bblly=161, bburx=552, bbury=736}
e)
\epsfig{figure=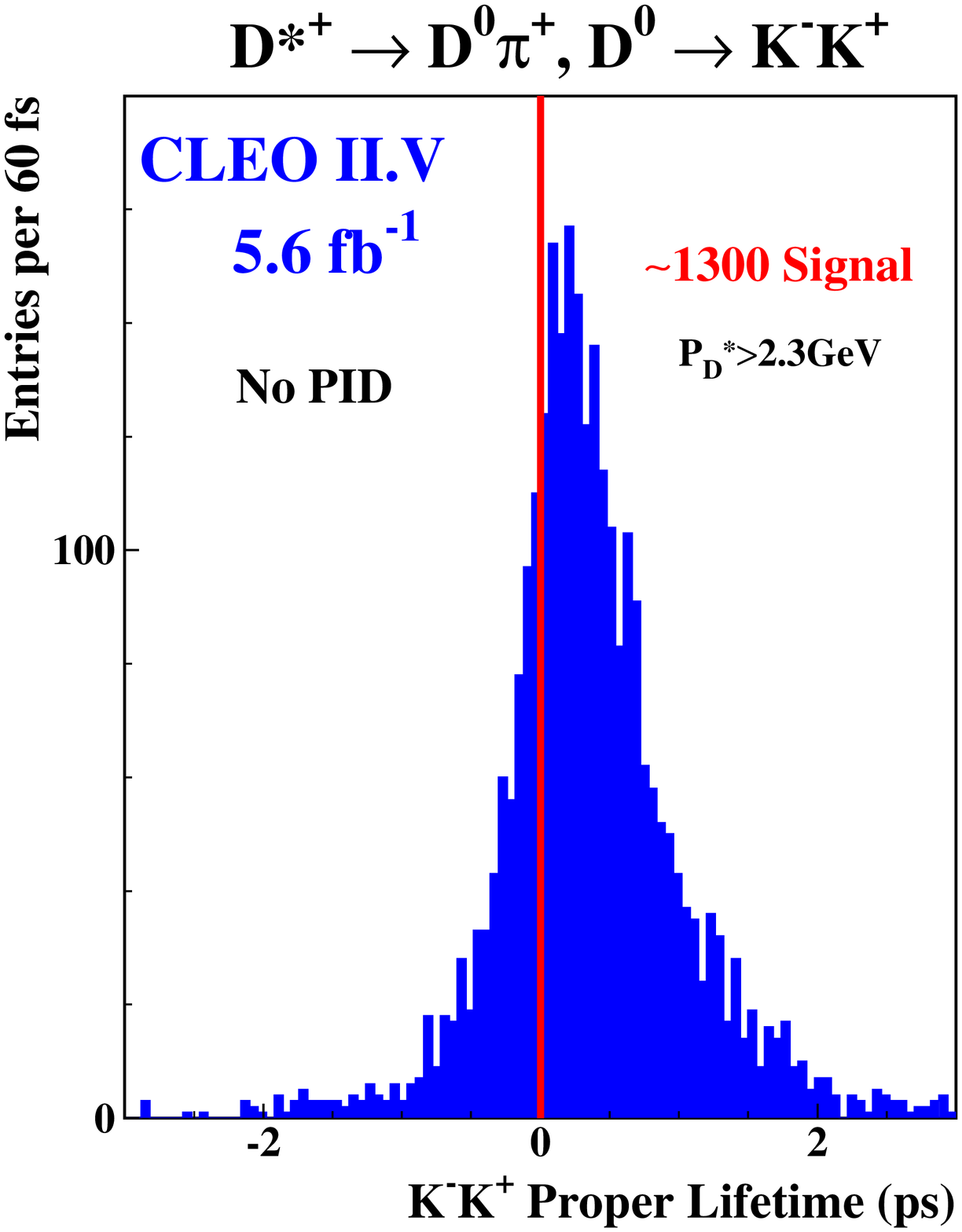,width=.24\textwidth,
height=1.in,clip=, bbllx=42, bblly=161, bburx=552, bbury=736}}
\caption{$CP^+$ Mass Distributions a) $\DZ \rightarrow K^+K^-$ b) $\DZ
\rightarrow \pi^+\pi^-$
with $D^{*\pm}$ tag. \ \ \ \ \ \ \ \ \
$CP^-$ Mass Distribution c) $\DZ \rightarrow \phi
K_{\rm S}$. \DZ \ decay time d) $\DZ \rightarrow K^+K^-$ e)
$\DZ \rightarrow \pi^+\pi^-$. \label{fig:CP}}
\end{figure}

The study of Cabibbo suppressed decays of the $\DZ$ to pairs
of pseudo-scalars provides two avenues into the study of
$\DZ\!-\!\DZB$ mixing.
First, for single Cabibbo suppressed
decays, the final states $\pi^+\pi^-$ and $K^+K^-$ (Fig.~\ref{fig:CP}a
and ~\ref{fig:CP}b) are common
to both the $\DZ$ and $\DZB$
, and so these final states provide innate sensitivity to mixing
.  Because these final states are
 also
$CP$ eigenstates, on-shell mixing of the $\DZ$ with the $\DZB$ can
change the exponential lifetime of the $\DZ$ as measured exclusively with
$\pi^+\pi^-$ and $K^+K^-$.  The shift in lifetime as measured with
$CP=+1$ final states, such as $\pi^+\pi^-$ and $K^+K^-$ (Fig.~\ref{fig:CP}d
and ~\ref{fig:CP}e), should
be equal in magnitude and opposite in sign to the lifetime shift
as measured with $CP=-1$ final states, such as $\rho^0K_{\rm S}$
and $\phi K_{\rm S}$ (Fig.~\ref{fig:CP}c).  
The $D^{*\pm}$ tag, used to identify the flavor of the decaying
$\DZ$ or $\DZB$, opens up the second avenue to mixing.  
The tag is essential to distinguish the  nominally double-Cabibbo 
suppressed decay (DCSD), $\DZ\!\to\!K^+\pi^-$, from the Cabibbo-favored
$\DZB\!\to\!K^+\pi^-$.  The time-integrated rate of 
$\DZ\!\to\!K^+\pi^-$ can then be used to limit the mixing process
$\DZ\!\to\!\DZB\to K^+\pi^-$. The proper time distribution for this
decay has three components - DCSD $\propto e^{-t}$, on-shell mixing
$\propto te^{-t}$ and off-shell mixing 
$\propto t^2e^{-t}$.
The contribution of DCSD is important to measure
because the smaller the DCSD contribution is, the
\emph{greater} the sensitivity to mixing.

\section{Formalism}

\begin{figure}[h]
\parbox{.4\textwidth}{a)}\parbox{.59\textwidth}{b)}
\parbox{.4\textwidth}{
\vskip .25cm
\epsfig{figure=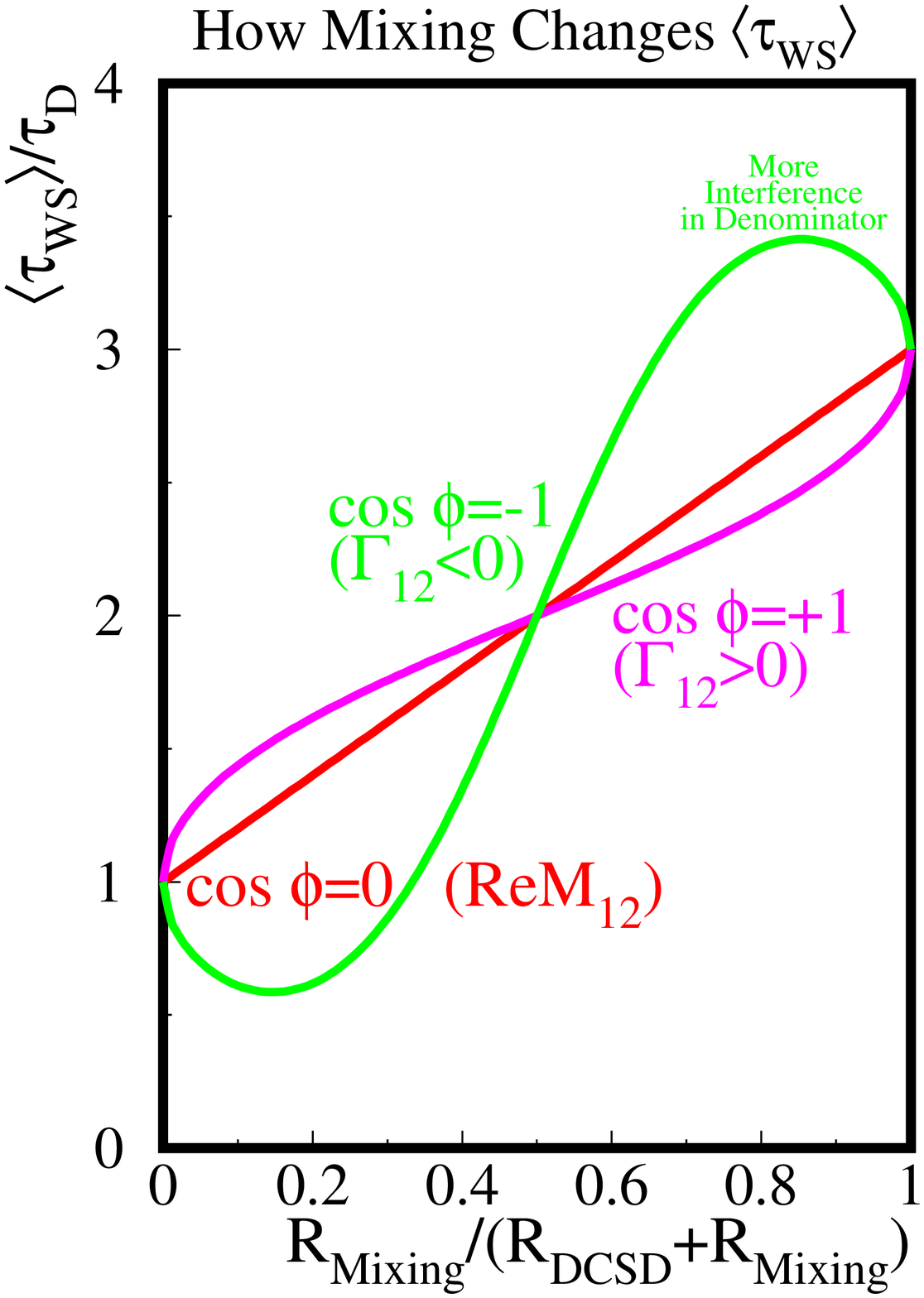,angle=-0.,width=.4\textwidth,height=2.0in,
clip=, bbllx=25, bblly=32, bburx=586, bbury=732}}
\parbox{.59\textwidth}{\epsfig{figure=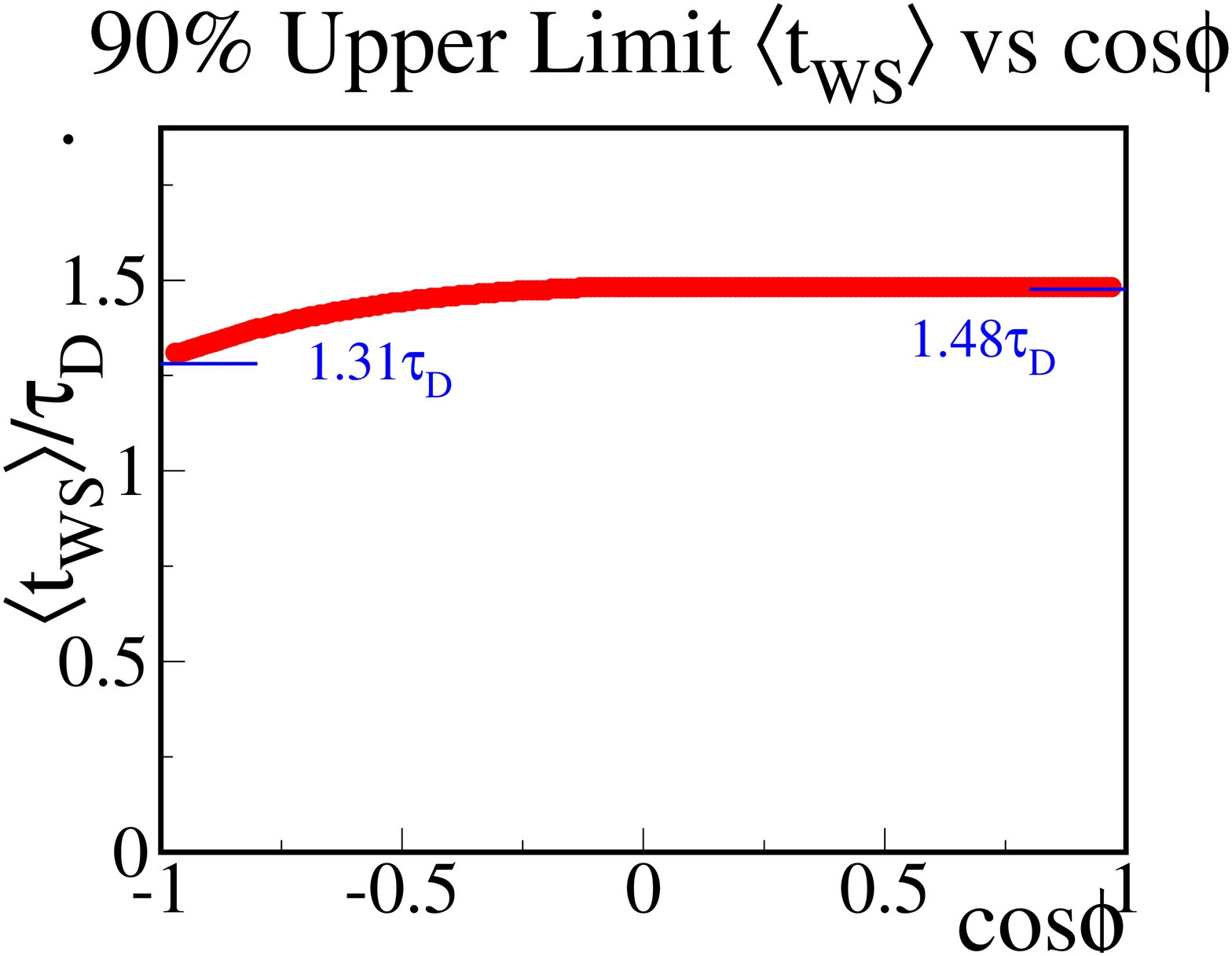,width=.59\textwidth,height=2.0in}}
\caption{a) $\langle t_{\rm WS}\rangle$ vs $R_{\rm Mix}/(R_{\rm DCSD}+R_{\rm Mix})$ 
b)$\langle t_{\rm WS} \rangle$ vs $\cos \phi$, $90\%$ C.L. Upper Limit.\label{fig:taulim}}
\end{figure}

Wrong-sign hadronic decays occur via DCSD or mixing. In the limit of small
mixing and no CP violation the decay time distribution depends on the
rates, $R_{\rm DCSD}$ and $R_{\rm Mix}$.
\begin{equation}
w(t)=(R_{\rm DCSD}+\sqrt{2R_{\rm DCSD}R_{\rm Mix}}\cos\phi\,t + \frac{1}{2}R_{\rm Mix}t^2)e^{-t}
\label{eq:wsdtau}
\end{equation}
where, in terms of the other usual parameters,
\begin{eqnarray}
	R_{\rm Mix} & = & \frac{1}{2}(x^2+y^2) \;\;\;\
	\phi  =  \tan^{-1}\left(-2\frac{\Delta M}{\Delta \Gamma} \right)
	+ \delta_s = \tan^{-1}\left(-{x\over y}\right) + \delta_s \\ \nonumber
\end{eqnarray}
The strong phase between $\DZ\to\WS$ \ and \ $\DZB \to \WS $ amplitudes,
$\delta_s$, is small by theoretical bias~\cite{browder95}.
The time-integrated wrong-sign rate is,
\begin{equation}
R_{\rm WS}=R_{\rm DCSD}+\sqrt{2R_{\rm DCSD}R_{\rm Mix}}\cos\phi+R_{\rm Mix},
\label{eq:wsrat}
\end{equation}
and the mean wrong-sign decay time is,
\begin{equation}
\langle t_{\rm WS}\rangle={{R_{\rm DCSD}+2\sqrt{2R_{\rm Mix}R_{\rm DCSD}}
\cos\phi+ 3R_{\rm Mix}}\over
    {R_{\rm DCSD}+\sqrt{2R_{\rm Mix}R_{\rm DCSD}}\cos\phi+R_{\rm Mix}}}
\label{eq:wstau}
\end{equation}
The behavior of $\langle t_{\rm WS}\rangle$ is shown as a function
of $R_{\rm Mix}/(R_{\rm DCSD}+R_{\rm Mix})$ in Fig.~\ref{fig:taulim}a for
the cases of $\cos\phi=\pm1$ and $\cos\phi=0$.  

\section{Wrong-Sign rate $R_{ws}$ and Mean Decay time $\langle t_{ws} \rangle$}
\begin{figure}[h]
\parbox{.32\textwidth}{a)}\parbox{.32\textwidth}{b)}\parbox{.32\textwidth}{c)}
\parbox{.32\textwidth}{\epsfig{figure=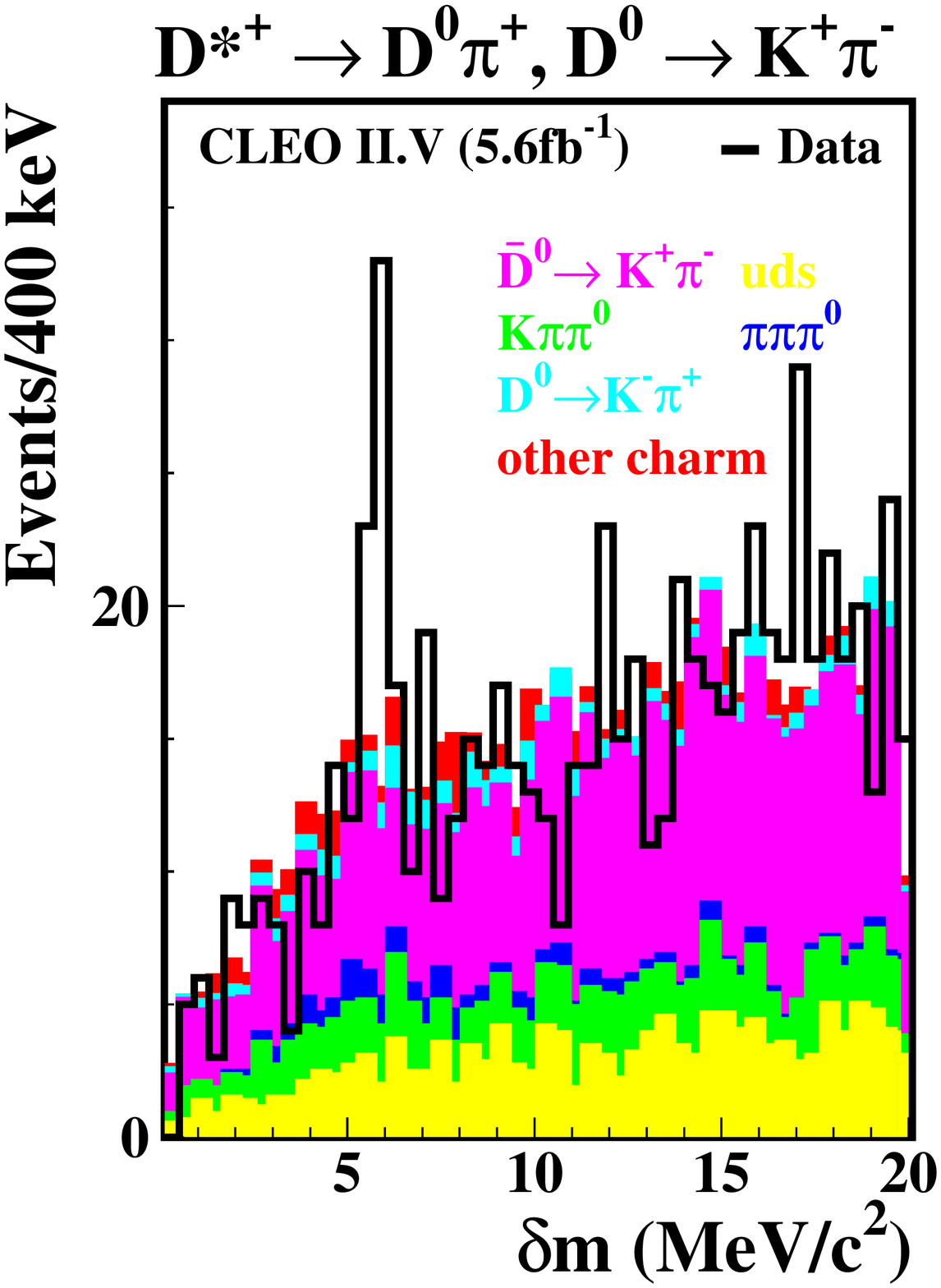,width=.32\textwidth,
height=2.5in,clip=, bbllx=62, bblly=70, bburx=561, bbury=732}}  \hfill
\parbox{.32\textwidth}{\epsfig{figure=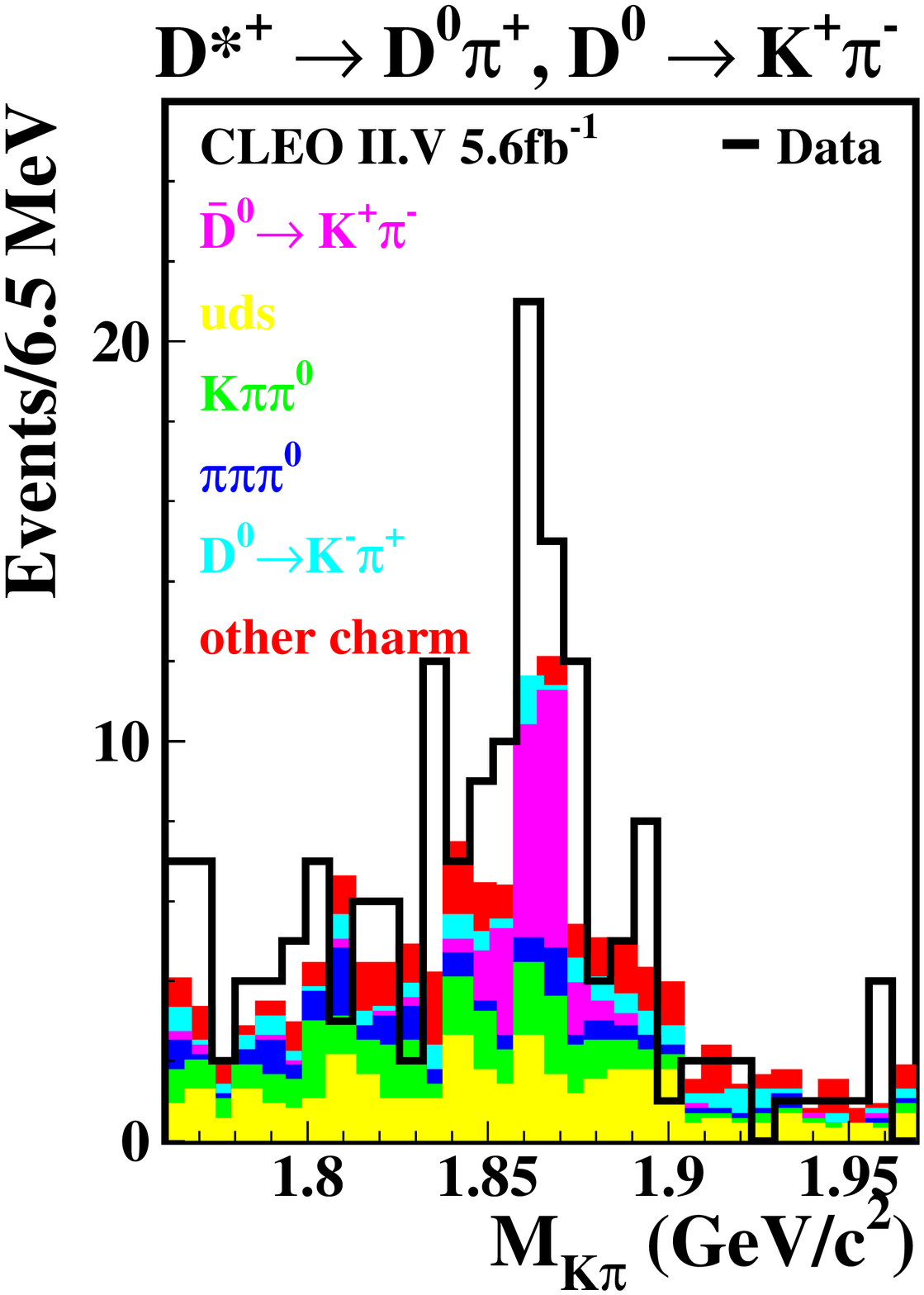,width=.32\textwidth,
height=2.5in,clip=, bbllx=62, bblly=70, bburx=561, bbury=732}}  \hfill
\parbox{.32\textwidth}{\epsfig{figure=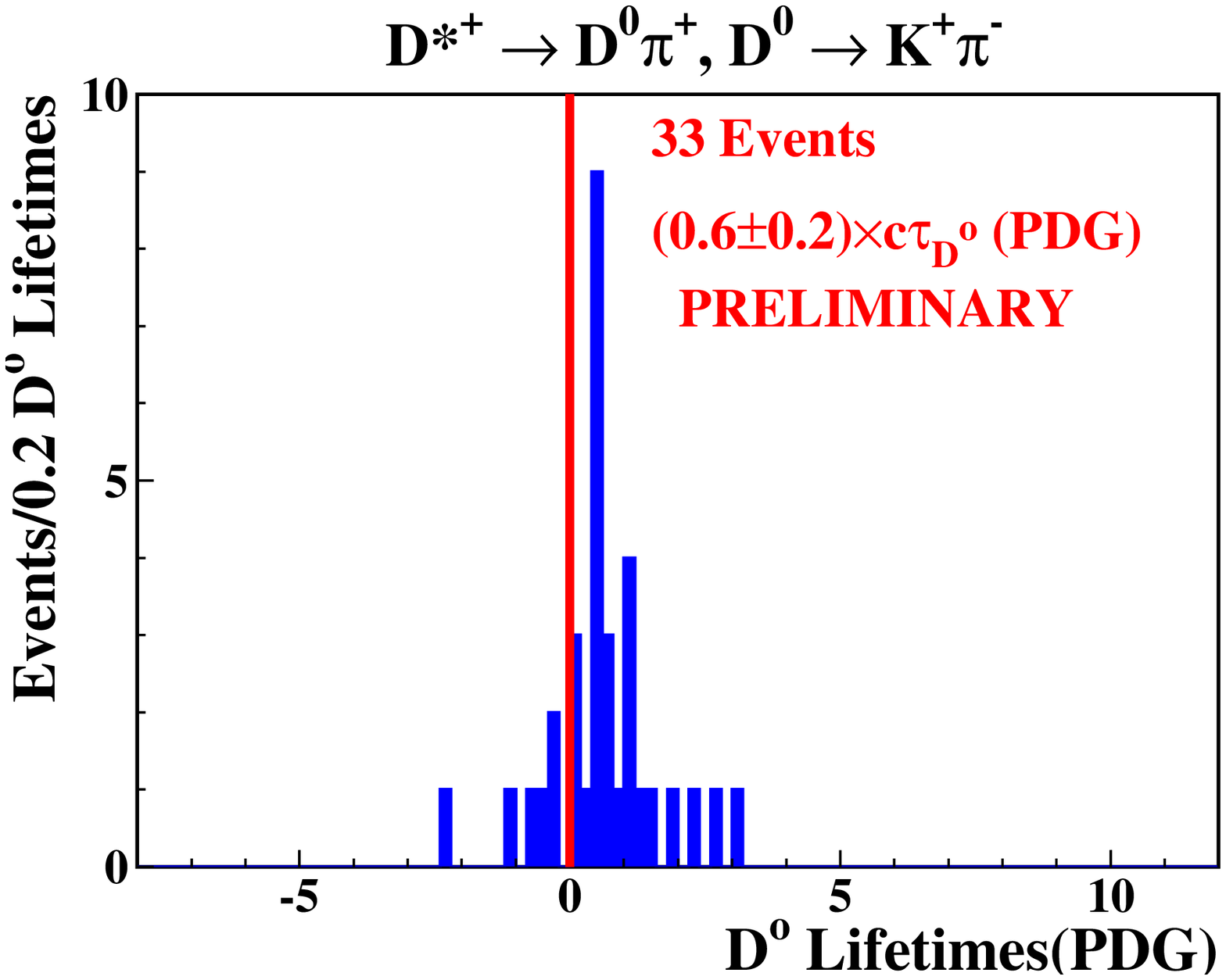,width=.32\textwidth,height=1.1in,
clip=, bbllx=14, bblly=208, bburx=540, bbury=594}
\parbox{.32\textwidth}{d)}
\epsfig{figure=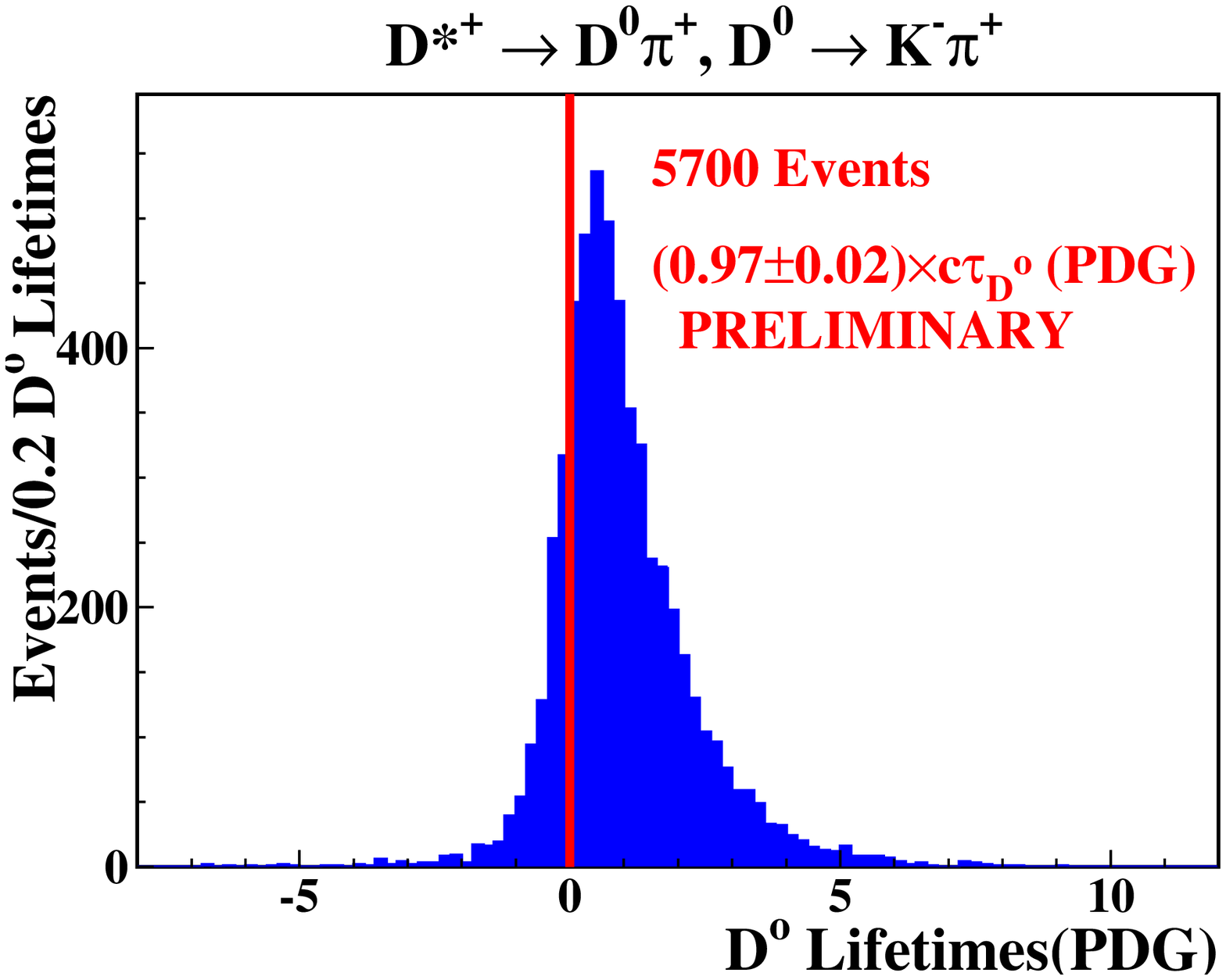,width=.32\textwidth,height=1.1in,
clip=, bbllx=14, bblly=208, bburx=540, bbury=594}}
\caption{Binned likelihood fit of a) $M_{K\pi\pi}-M_{K\pi}-M_{\pi}$
 vs b) $M_{K\pi}$ 
determines composition of background. $t_{ws}$ for c) $\DZ\to\WS$ d) $\DZ\to\KPI$.\label{fig:rwstws}}
\end{figure}

A binned maximum likelihood fit of the MC-generated
background components to the two dimensional data on the
$M_{K\pi\pi}-M_{K\pi}$ vs. $M_{K\pi}$ plane\cite{dpf99} determines $R_{ws}$.
\begin{equation}
R_{\rm WS}  = {\Gamma(\DZ\to\WS)\over\Gamma(\DZ\to\KPI)}\,=\,0.0031 \pm .0009(stat) \pm .0007(syst)
\label{eq:rws}
\end{equation}
The fit also yields a breakdown of the background
event content in Fig.~\ref{fig:rwstws}a and ~\ref{fig:rwstws}b. 
The mean Wrong-sign decay time can be determined from Fig.~\ref{fig:rwstws}c
using the mean decay time for $\DZB$ \ and
$uds$ backgrounds of $\tau=1$ and $\tau=0$, respectively, combined with the
background composition, we evaluate:
\begin{equation}
\langle t_{\rm WS}\rangle  = (0.65\pm0.40)\;\;\;(\times \tau_{D^0})
\label{eq:tws}
\end{equation}

Proper renormalization to the physical regions of $t_{ws}$ (Fig.~\ref{fig:taulim}a) is required.
The $90\%$ C.L. Upper Limit on $\langle t_{\rm WS}\rangle$ vs $\cos \phi$
is shown in Fig.~\ref{fig:taulim}b.  
We obtain limits in the two dimensional space of
$R_{\rm Mix}$ vs. $R_{\rm DCSD}$ from the rate
of Wrong Sign decay, and the mean $\langle t_{\rm WS}\rangle$.

\section{Previous $\DZ\!-\!\DZB$ \ Mixing Limits}
Three groups have reported non-zero measurements of $R_{\rm WS}$
all with analysis evaluated for the case $\cos\phi=0$, and with
neglect of CP violation:
\begin{itemize}
\item CLEO-II\cite{paper_liu}, equivalent to 
      $R_{\rm WS}=R_{\rm DCSD}+R_{\rm Mix}=(0.77\pm0.35)\%$.
\item E791\cite{mix_e791_had}, where $R_{\rm DCSD}=(0.68\pm0.35)\%$,
      and $R_{\rm Mix}=(0.21\pm0.09)\%$, where, for $R_{\rm Mix}$,
      $\DZ\to K^+\pi^-\pi^+\pi^-$ contribute in addition to
      $\DZ\to K^+\pi^-$; no report of a non-zero
      $R_{\rm Mix}$ was made.
\item Aleph\cite{mix_aleph}, where $R_{\rm DCSD}=(1.84\pm0.68)\%$, and
      an upper limit of $R_{\rm Mix}<0.92\%$ is obtained, at
      $95\%$ C.L.
\end{itemize}
Additionally, 
there are two other relevant limits on $R_{\rm WS}$.
The E691 collaboration\cite{mix_e691} limited $R_{\rm Mix}<0.37\%$,
at 90\% C.L., where again $\DZ\to K^+\pi^-\pi^+\pi^-$ 
contribute in addition to $\DZ\to K^+\pi^-$,
and $R_{\rm DCSD}<1.5\%$ at 90\% C.L.
The E791\cite{mix_e791_lep} collaboration sought 
$\DZ\!\to\!K^+\ell^-\overline{\nu}_{\ell}$,
and limited $R_{\rm Mix}<0.5\%$.  
The regions allowed by the above work, in the $R_{\rm Mix}$ 
vs. $R_{\rm DSCD}$ plane, for
$\cos \phi=0$, are shown in Fig.~\ref{fig:mixlim}a.
\section{CLEO-II.V Charm Mixing Limits}
\begin{figure}[t]
\parbox{.32\textwidth}{a)}\parbox{.32\textwidth}{b)}\parbox{.32\textwidth}{c)}
\parbox{.32\textwidth}{\epsfig{figure=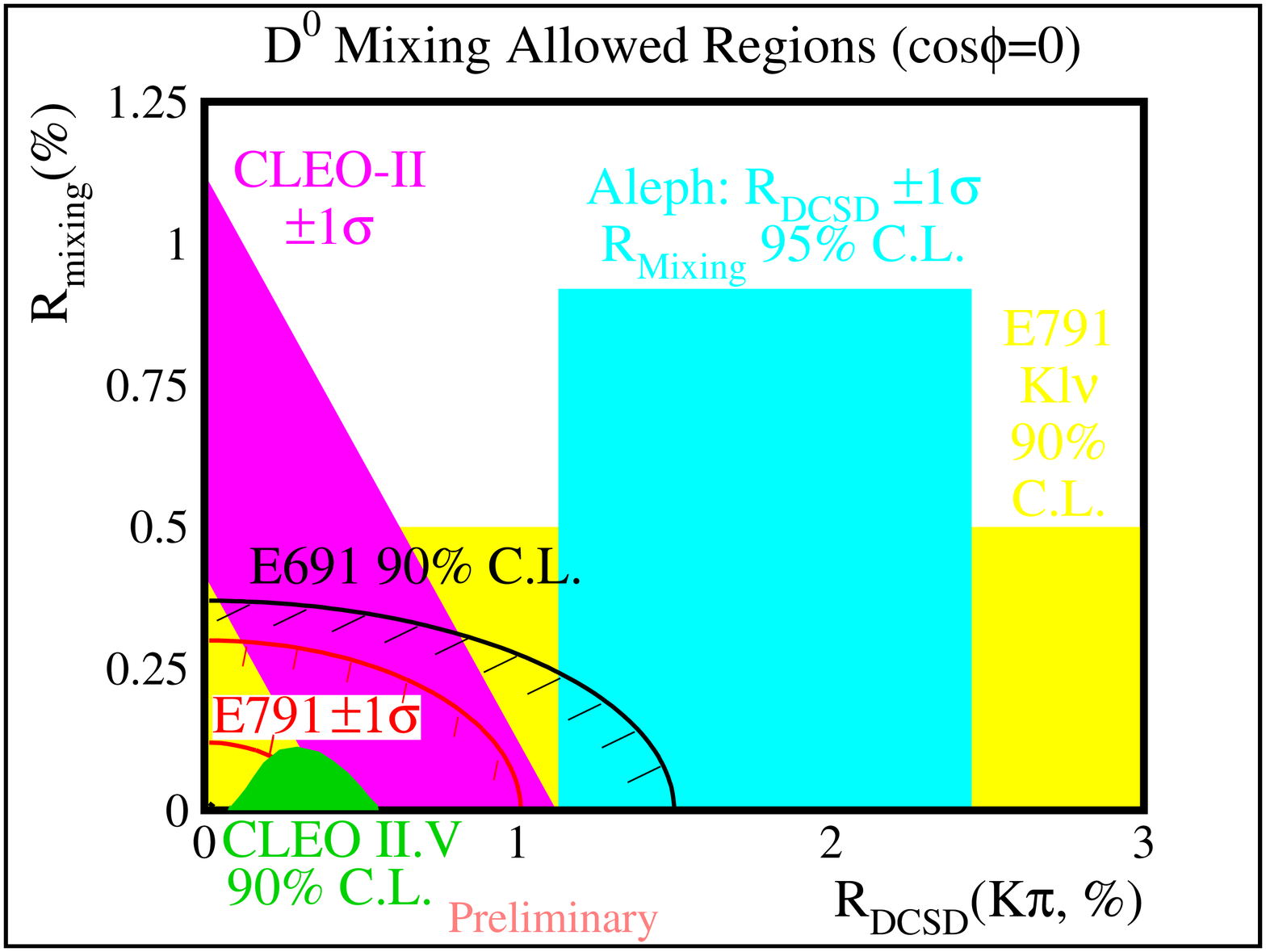,width=.4\textwidth,height=1.5in,angle=-90}}
\parbox{.32\textwidth}{\epsfig{figure=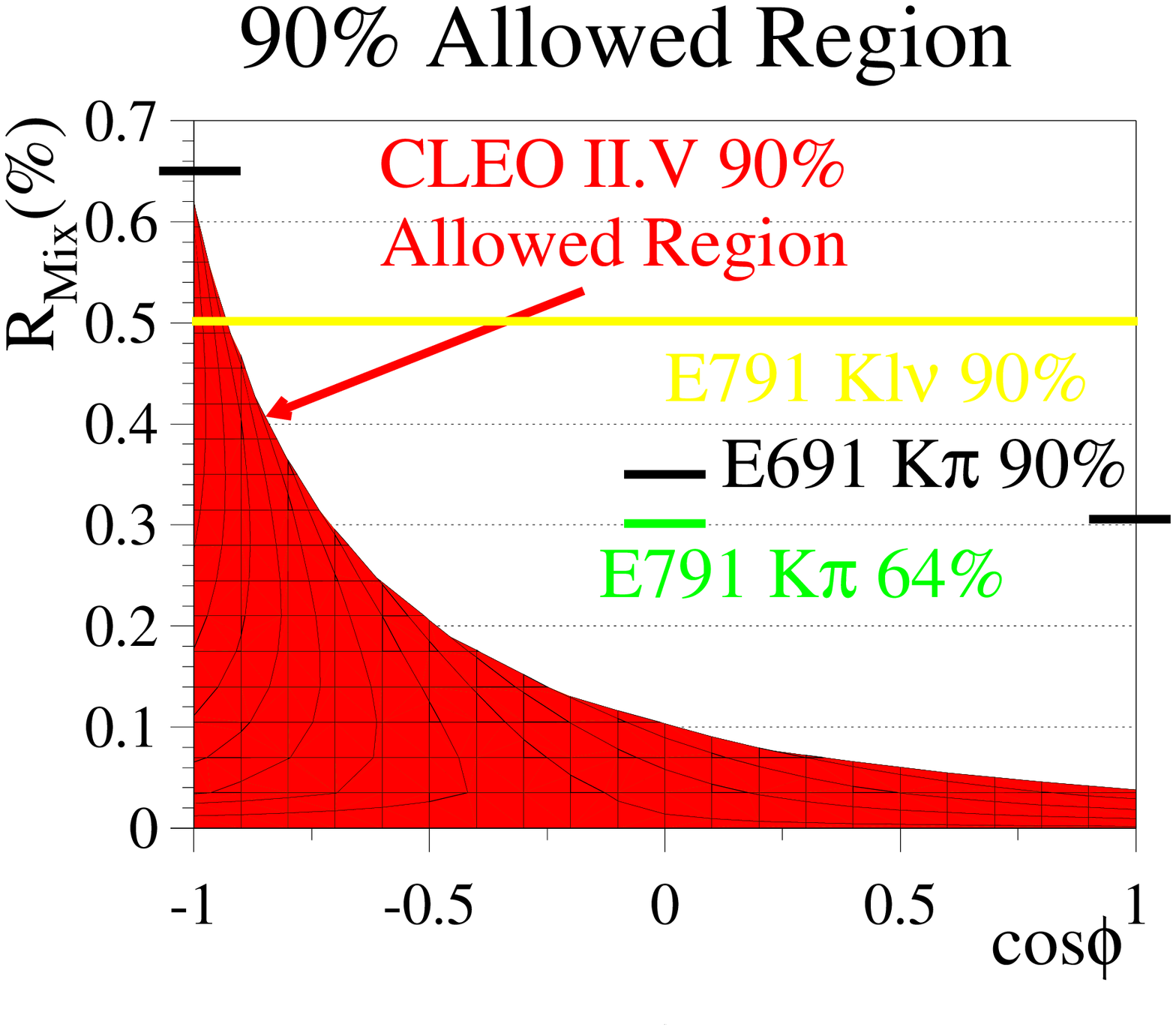,width=.32\textwidth,height=2.0in
,clip=, bbllx=8, bblly=190, bburx=540, bbury=580}}
\parbox{.32\textwidth}{\epsfig{figure=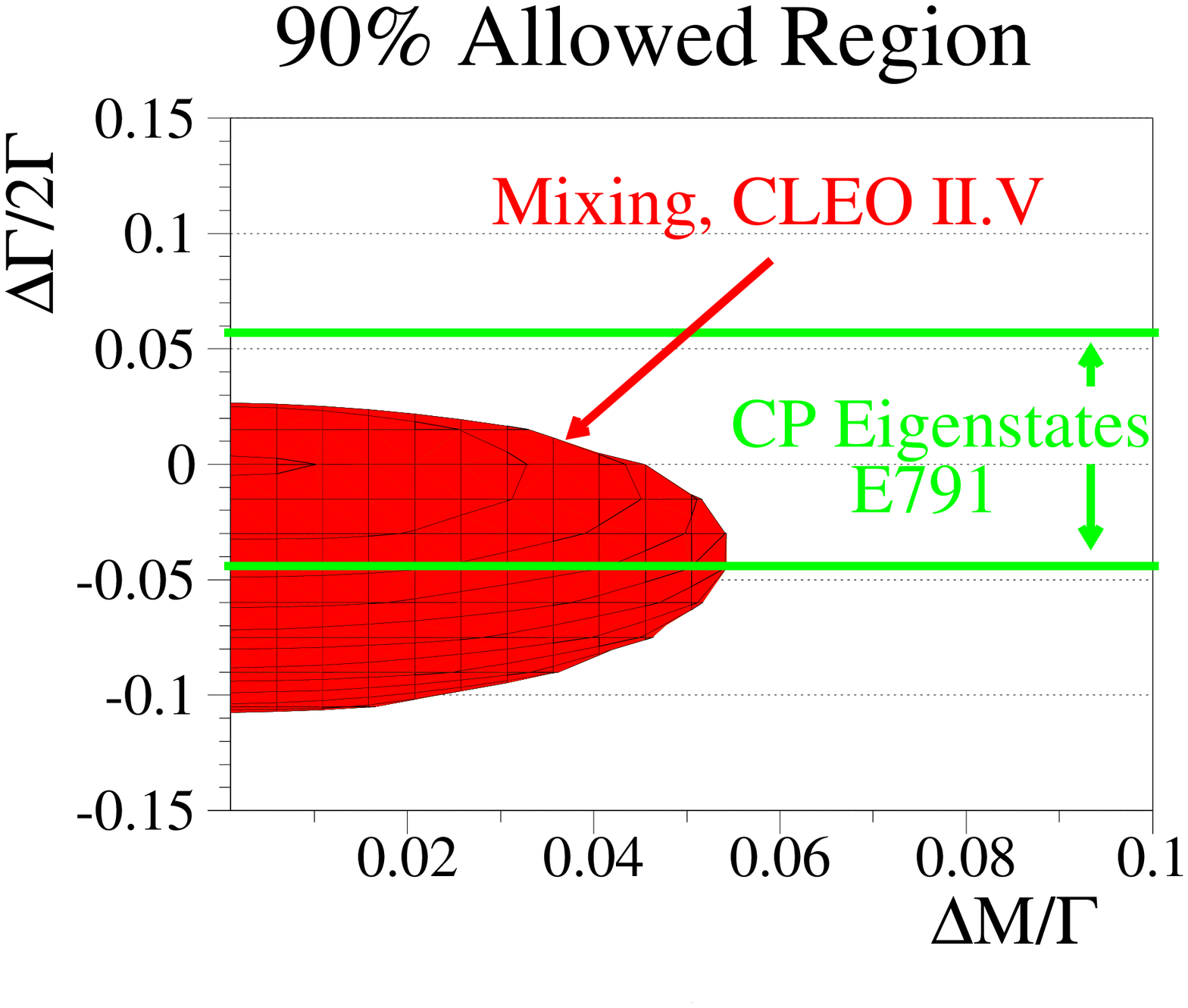,width=.32\textwidth,height=2.0in
,clip=, bbllx=4, bblly=190, bburx=540, bbury=580}}
\caption{a) World mixing limits ($\cos \phi = 0$). CLEO-II.V $90\%$ Mixing limits
obtained from $\DZ\to\WS$ \ b) $R_{Mix}$ $vs$ $\cos \phi$. c) $x$ $vs$ $y$. 
\label{fig:mixlim}}
\end{figure}

{\label{sec:mixlim}}
The limits on $\DZ\!-\!\DZB$ determined from $\DZ\to\WS$ \ with 5.6$fb^{-1}$ of
CLEO-II.V data are shown in Fig.~\ref{fig:mixlim}b-c 
and in column 1 of table~\ref{tbl:mixtab}.
Combining with $\DZ \rightarrow CP^+$ analysis from
E791\cite{mix_e791_CP} improves limits on $R_{Mix}$ and $y$ 
(table~\ref{tbl:mixtab},column 2). The
CLEO-II.V sensitivity (9.1$fb^{-1}$) combining
$\DZ\to\WS,K^+\pi^-\pi^0,K^+\pi^-\pi^+\pi^-$ and $\DZ \rightarrow CP$
analyses is listed in column 3. A factor of 2-5 (3-10) improvement in
precision is obtained over the PDG\cite{RPP98} with 5.6$fb^{-1}$
(9.1$fb^{-1}$). It is noteworthy that the
CLEO II.V limit for $x$ $\sim\tan^2\theta_{\rm Cabibbo}$, is more or less the
largest level that $\DZ\!-\!\DZB$ mixing can be in the Standard Model. 
\begin{table}[h]
\caption{Current limit on $\DZ\!-\!\DZB$ \ Mixing Limits and projected CLEO-II.V sensitivity.}
\label{tbl:mixtab}
\begin{center}
\begin{tabular}{|l|c|c|c|c|} \hline
 & CLEO-II.V & CLEO-II.V & CLEO-II.V & RPP98 \\
 & ($5.6fb^{-1}$) & +E791 & (Complete) & \\ \hline
$x$& $\hphantom{-.108<}\left|x\right|<.054$ &  $\hphantom{-.042<}\left|x\right|<.054$ &  $\left|x\right|<.03$ &  $\left|x\right|<.096$\\
$y$ & $-.108<\;y\;<.027$ & $-.042<\;y\;<.027$ & $\left|y\right|<.01$ & $\left|y\right|<.10$ \\
$R_{ws}$ & $.31\pm.09\%$ &  $.31\pm.09\%$ &  $\pm.05\%$ &  $.72\pm.25\%$\\ 
$R_{Mix}$ & $<1.1\%$ &  $<0.25\%$ &  $<0.05\%$&  $<0.5\%$\\ \hline
\end{tabular}
\end{center}
\end{table}
\section*{Acknowledgments}
We gratefully acknowledge the effort of the CESR staff in providing us with
excellent luminosity and running conditions.
This work was supported by
the National Science Foundation,
the U.S. Department of Energy,
Research Corporation,
the Natural Sciences and Engineering Research Council of Canada,
the A.P. Sloan Foundation,
the Swiss National Science Foundation,
and the Alexander von Humboldt Stiftung.

\end{document}